\documentclass[conference]{IEEEtran}
\IEEEoverridecommandlockouts

\usepackage{cite}
\def\BibTeX{{\rm B\kern-.05em{\sc i\kern-.025em b}\kern-.08em
    T\kern-.1667em\lower.7ex\hbox{E}\kern-.125emX}}

\usepackage[utf8]{inputenc}
\usepackage[T1]{fontenc}
\usepackage{amsmath,amssymb,amsfonts}
\usepackage{graphicx}
\usepackage{textcomp}
\usepackage{booktabs}
\usepackage[caption=false,font=footnotesize]{subfig}
\usepackage{adjustbox}
\usepackage{pgfplots}
\usepackage{xcolor}
\pgfplotsset{compat=1.18}
\usepackage{newtxtext,newtxmath} 
\usepackage{tabularx}

\usepackage{listings}
\usepackage{needspace}
\usepackage{mathtools}

\usepackage{letltxmacro}
\LetLtxMacro{\oldthebibliography}{\thebibliography}
\renewcommand{\thebibliography}[1]{%
  \oldthebibliography{#1}%
  \setlength{\itemsep}{0pt}%
}

\definecolor{mygreen}{rgb}{0,0.6,0}
\definecolor{mygray}{rgb}{0.5,0.5,0.5}
\definecolor{mymauve}{rgb}{0.58,0,0.82}

\usepackage{xurl} 

\newcommand{\vect}[1]{\boldsymbol{#1}}
\newcommand{\mat}[1]{\mathbf{#1}}
\usepackage{hyperref}
\hypersetup{hidelinks}

\begin{document}

\title{Decentralized Local Voltage Control for Active Distribution Networks
\thanks{This work was co-financed by Funda\c{c}\~{a}o para a Ci\^{e}ncia e a Tecnologia through the Carnegie Mellon Portugal Program under the fellowship FCT 2025.00027.PRT.}
}

\author{%
  \IEEEauthorblockN{Diana Vieira Fernandes}
  \IEEEauthorblockA{\textit{Dept.\ of Engineering \& Public Policy}\\
                    \textit{Carnegie Mellon University, USA}\\
                    \textit{IN+/LARSyS, Instituto Superior Técnico}\\
                    \textit{Universidade de Lisboa, Portugal}\\
                    \small\textit{dianaimf@andrew.cmu.edu}}
  \and
  \IEEEauthorblockN{Soummya Kar}
  \IEEEauthorblockA{\textit{Dept.\ of Electrical \& Computer Engineering}\\
                    \textit{Carnegie Mellon University, USA}\\
                    \small\textit{soummyak@andrew.cmu.edu}}
  \and
  \IEEEauthorblockN{Carlos Santos Silva}
  \IEEEauthorblockA{\textit{IN+/LARSyS}\\
                    \textit{Instituto Superior Técnico}\\
                    \textit{Universidade de Lisboa, Portugal}\\
                    \small\textit{carlos.santos.silva@tecnico.ulisboa.pt}}}

\maketitle
\begin{abstract}
Distribution networks face challenges from the increasing deployment of Distributed Energy Resources (DERs) and the emergence of bidirectional power flows. We propose a decentralized Volt/VAr control method based on a saddle-point reformulation and consensus+innovation (C+I) updates. Each agent at a controllable bus computes and enforces its own set‑points using only neighbor communication. Our method embeds passive buses directly, preserves network physics through a linearized Jacobian model, and avoids any supervisory nodes. Simulation results on a modified CIGRE low-voltage network show voltage stability improvement within operational limits, indicating the viability of a fully decentralized (edge‑based) Volt/VAr control solution.
\end{abstract}

\section{Introduction}
As electric power systems evolve towards decentralized and distributed paradigms, the integration of Distributed Energy Resources (DERs) has exposed distribution networks to localized voltage violations and congestion, challenging traditional voltage regulation methods. Passive control devices, which were previously sufficient for managing voltage profiles in radial networks, are increasingly inadequate in the face of intermittent renewable generation and bidirectional power flows \cite{bollen_voltage_2005}. Newer strategies -- ranging from inverter-based reactive power support to coordinated active/reactive power dispatch -- have shown promise in enhancing voltage control \cite{wu_distributed_2017}, particularly under the constraints of unbalanced or meshed networks \cite{fusco_decentralized_2021,srivastava_voltage_2023}. Control architectures have diversified into centralized, distributed, and local schemes. Centralized solutions offer global optimality, but suffer from scalability and latency issues in real-time applications \cite{hasan_centralized_2024}. Distributed and hierarchical control methods -- often leveraging consensus-based protocols and optimization decomposition -- distribute computation among agents, improving scalability and robustness in systems with extensive DERs \cite{tang_distributed_2021, pierrou_decentralized_2024}. Local controllers, while faster and communication-free, lack coordination and global awareness, making them suitable primarily for simple or constrained scenarios \cite{zhou_reverse_2021,andren_stability_2015}. The choice of power flow modeling significantly influences control strategies. The Bus Injection Model (BIM) and Branch-Flow Model (DistFlow) each present trade-offs: BIM supports unbalanced multiphase modeling \cite{bollen_voltage_2005}, while DistFlow simplifies computation in radial networks but performs poorly under reverse flows \cite{gan_convex_2014, kekatos_voltage_2016}. Various convex relaxations -- such as semidefinite programming (SDP) and second order cone programming (SOCP) -- aim to overcome nonlinearity in AC power flow \cite{gan_convex_2014, kim_exact_2003}. Recent works include coordination strategies involving Virtual Power Plants (VPPs) \cite{dallanese_optimal_2018} or microgrids \cite{gray_distributed_2023}.
\section{Contributions}
We propose a novel fully decentralized agent-based framework for Volt/VAr control that integrates network constraints and supports multiple, concurrent autonomous agents with voltage control capacity (controllability). This work addresses aspects not covered comprehensively in existing literature, i.e., by enabling system-wide coordination without relying on any centralized entity. Our key contributions include:

\begin{itemize}
    \item \textbf{Autonomous Corrective Control:} Autonomous agents compute optimal set-points and execute corrective actions to regulate voltages within limits without external intervention. Passive nodes are embedded directly in the constraint set, so the method respects voltage and current limits even where no control authority exists.
    \item \textbf{Fully Decentralized agent-based control architecture:} We apply a saddle-point reformulation to enable true decentralized control, distributing the system-wide optimization problem across agents at controllable buses  -- entirely to the network edge -- and merging the autonomy of local decision making with network‑level feasibility. Unlike hierarchical or clustered decentralized schemes surveyed \cite{srivastava_voltage_2023, fusco_decentralized_2021}, our approach operates without supervisory nodes. System-wide coordination is achieved through a combination of a consensus+innovation (C+I) mechanism for tracking a shared global control estimate and local dual variable updates that enforce linearized power balance constraints at each bus. Each DER source relies solely on locally measured voltages and injections, augmented with a consensus signal, removing the need for global state estimation.
    \item \textbf{Integrated Network Physics:} Explicitly incorporates detailed network constraints derived from linearized power flow, including modeling of uncontrollable passive buses directly within the problem constraints. The optimization problem avoids using simplified models, e.g., the DistFlow equations, which assume unidirectional power flow (using a directed graph structure), to ensure accuracy in networks with significant distributed generation.
    \item \textbf{Distributed Policy Synthesis:} We propose a novel use of consensus for the peer-to-peer synthesis of a global control policy vector. It directly embeds the coordination into the control synthesis loop. This allows each agent to see the system-wide consequences of its proposed actions (via the shared vector \(\vect{u}\)) and adjust them to be compliant with the overall grid physics, which are enforced by the separate primal-dual mechanism.
\end{itemize}

\section{Preliminaries}

\subsection{Static Linearized Model}

Linearizing the full AC power flow equations around a nominal operating point \((V^0,\theta^0)\) yields:
\begin{align}
P_i &= P_i^0 + \sum_{j\in\mathcal{N}} H_{ij}\,(\theta_j - \theta_j^0) + \sum_{j\in\mathcal{N}} N_{ij}\,(V_j - V_j^0), \label{eq:linearizedP}\\[1mm]
Q_i &= Q_i^0 + \sum_{j\in\mathcal{N}} K_{ij}\,(\theta_j - \theta_j^0) + \sum_{j\in\mathcal{N}} L_{ij}\,(V_j - V_j^0), \label{eq:linearizedQ}
\end{align}
where \(H_{ij}\), \(N_{ij}\), \(K_{ij}\), and \(L_{ij}\) are sensitivity coefficients derived from the Jacobian of the power flow equations.

Defining the angle and voltage deviations (\(\Delta \theta_j = \theta_j - \theta_j^0,\quad \Delta V_j = V_j - V_j^0\)), we can write the above equations in a compact form as:
\begin{equation}
    \begin{pmatrix} \Delta P \\[1mm] \Delta Q \end{pmatrix} =
\underbrace{\begin{pmatrix} H & N \\[1mm] K & L \end{pmatrix}}_{J}
\begin{pmatrix} \Delta \theta \\[1mm] \Delta V \end{pmatrix},
\end{equation}
with \(\Delta P = P - P^0\) and \(\Delta Q = Q - Q^0\). 
Assuming the reduced Jacobian (with the slack bus removed) is nonsingular, we use \(\mat{J}^{-1}\) to mean the zero-padded inverse (slack rows/cols are zero)
\footnote{Designating one bus as slack and removing its two rows and columns yields the full‑rank reduced matrix of size \(2(n-1)\times 2(n-1)\), where \(n = |\mathcal{N}|\). The original \(2n \times 2n\) Jacobian is recovered by padding zeros that lock the slack variables at their reference values. Let \(i_{\mathrm{ref}}\in\mathcal N\) be the slack bus. For any reduced \(2(n-1)\times 2(n-1)\)  matrix $\mat{J}_{\mathrm{red}}$ obtained by removing the slack–bus row and column, we write \(\mat{J}_{\text{full}} := \text{pad}\bigl(\mat{J}_{\mathrm{red}}\bigr)\in\mathbb{R}^{2n\times 2n}\), meaning that a row and column of zeros are inserted in the \(i_{\mathrm{ref}}\) position. Henceforth, Jacobian-sized objects are \(2n\times 2n\) after padding, while bus-indexed vectors/matrices are \(n\)-dimensional; entries at the slack bus are fixed (zero deviations).}.
We denote its inverse by \( \mat{J}^{-1} = \begin{pmatrix} \tilde{\mat{H}} & \tilde{\mat{N}} \\[1mm] \tilde{\mat{K}} & \tilde{\mat{L}} \end{pmatrix}\), where \(\tilde{\mat{H}} = \frac{\partial \Delta \theta}{\partial P}\) (angle response to active power), \(\tilde{\mat{N}} = \frac{\partial \Delta \theta}{\partial Q}\) (angle response to reactive power), \(\tilde{\mat{K}} = \frac{\partial \Delta V}{\partial P}\) (voltage response to active power), and \(\tilde{\mat{L}} = \frac{\partial \Delta V}{\partial Q}\) (voltage response to reactive power). Then, for each bus \(i\), the deviations are given by:
\begin{align}
\Delta \theta_i &= \sum_{j\in\mathcal{N}} \tilde{\mat{H}}_{ij}\,(P_j-P_j^0) + \sum_{j\in\mathcal{N}} \tilde{\mat{N}}_{ij}\,(Q_j-Q_j^0), \label{eq:theta_deviation}\\[1mm]
\Delta V_i &= \sum_{j\in\mathcal{N}} \tilde{\mat{K}}_{ij}\,(P_j-P_j^0) + \sum_{j\in\mathcal{N}} \tilde{\mat{L}}_{ij}\,(Q_j-Q_j^0). \label{eq:V_deviation}
\end{align}

The terms  (\(\mat{H}, \mat{N}, \mat{K}, \mat{L})\) refer to the submatrices of the forward Jacobian matrix \(\mat{J}\). The tilde notation \((\tilde{\mat{H}}, \tilde{\mat{N}}, \tilde{\mat{K}}, \tilde{\mat{L}})\) denotes the corresponding submatrices of the inverse Jacobian, \(\mat{J}^{-1}\). These are elements in \(\mathbb{R}^{n \times n}\), where \(n = |\mathcal{N}|\) is the total number of buses. 
\subsection{State-Space Model}
To facilitate control design, we reformulate the linearized power flow equations into a state-space representation. Let the state vector be $\vect{x} = [\Delta \vect{\theta}^\top, \Delta \vect{V}^\top]^\top \in \mathbb{R}^{2|\mathcal{N}|}$. The state-space equation is:
\begin{equation}
\vect{x} = \mat{J}_{\text{ctrl}}^{-1} \vect{u} + \vect{b},
\label{eq:state_space}
\end{equation}
where: 
\begin{itemize}
  \item $\mat{J}_{\text{ctrl}}^{-1} \in \mathbb{R}^{2|\mathcal{N}|\times 2|\mathcal{N}_{\text{ctrl}}|}$ maps controllable injection deviations to state deviations (columns of $\mat J^{-1}$ restricted to controllable buses),
  \item $\vect{b} \in \mathbb{R}^{2|\mathcal{N}|}$ is a constant offset determined by the linearization point and non-controlled deviations; it is treated as known and fixed during the update.
\end{itemize}
The compact control vector $\vect{u} \in \mathbb{R}^{2|\mathcal{N}_{\text{ctrl}}|}$ stacks the local effective injection vectors $\hat{\vect{u}}_i$ from each controllable bus $i \in \mathcal{N}_{\text{ctrl}}$:
\begin{equation}
\vect{u} =
\begin{bmatrix}
P_{\text{ctrl},c_1} z_{c_1} \\[1pt] Q_{\text{ctrl},c_1} z_{c_1} \\
\vdots \\ 
P_{\text{ctrl},c_{n_c}} z_{c_{n_c}} \\[1pt] Q_{\text{ctrl},c_{n_c}} z_{c_{n_c}}
\end{bmatrix}
\quad\text{with}\quad
\hat{\vect{u}}_i = \begin{bmatrix} P_{\text{ctrl},i}\,z_i \\ Q_{\text{ctrl},i}\,z_i \end{bmatrix}.\;\;
\label{eq:u_and_u_hat_def}
\end{equation}
To relate local control actions to the global control vector $\vect{u}$, we define an ordered set of controllable buses $\{c_1, \dots, c_{n_c}\}$ where $n_c=|\mathcal{N}_{\text{ctrl}}|$. For each bus $i \in \mathcal{N}_{\text{ctrl}}$, let $p(i)$ be its index in this set. The corresponding selector matrix $\mat{E}_i \in \mathbb{R}^{2|\mathcal{N}_{\text{ctrl}}| \times 2}$
is~\eqref{eq:E_i}:
\begin{equation}
\mat{E}_i = 
  \begin{bmatrix}
    \mat{0}_{2(p(i)-1),\,2} \\
    \mat{I}_{2} \\
    \mat{0}_{2|\mathcal{N}_{\text{ctrl}}|-2p(i),\,2}
  \end{bmatrix}
\label{eq:E_i}  
\end{equation}
such that \(\vect{u} = \sum_{i \in \mathcal{N}_{\text{ctrl}}} \mat{E}_i \vect{\hat{u}}_i\). This matrix is essential for the consensus mechanism, allowing each agent to inject its local information into the shared global estimate. 
The structure of this mapping is shown explicitly in \eqref{eq:full_block}:
\begin{equation}
\resizebox{\columnwidth}{!}{$
\underbrace{
\begin{bmatrix}
\Delta \theta_1 \\
\vdots \\
\Delta \theta_{|\mathcal N|} \\
\Delta V_1 \\
\vdots \\
\Delta V_{|\mathcal N|}
\end{bmatrix}
}_{\vect{x} \in \mathbb{R}^{2|\mathcal N|}}
=
\underbrace{
\left[
\begin{array}{@{}c@{\;}c@{\;}c@{\;}c@{\;}c@{\;}c@{}}
\tilde{\mat{H}}_{1,c_1} & \tilde{\mat{N}}_{1,c_1} & \cdots & \tilde{\mat{H}}_{1,c_{|\mathcal N_{\text{ctrl}}|}} & \tilde{\mat{N}}_{1,c_{|\mathcal N_{\text{ctrl}}|}}\\
  \vdots & \ddots & \vdots & \vdots & \ddots & \vdots\\
\tilde{\mat{H}}_{|\mathcal N|,c_1} & \tilde{\mat{N}}_{|\mathcal N|,c_1} & \cdots & \tilde{\mat{H}}_{|\mathcal N|,c_{|\mathcal N_{\text{ctrl}}|}} & \tilde{\mat{N}}_{|\mathcal N|,c_{|\mathcal N_{\text{ctrl}}|}}\\[2mm]
\tilde{\mat{K}}_{1,c_1} & \tilde{\mat{L}}_{1,c_1} & \cdots & \tilde{\mat{K}}_{1,c_{|\mathcal N_{\text{ctrl}}|}} & \tilde{\mat{L}}_{1,c_{|\mathcal N_{\text{ctrl}}|}}\\
  \vdots & \ddots & \vdots & \vdots & \ddots & \vdots\\
\tilde{\mat{K}}_{|\mathcal N|,c_1} & \tilde{\mat{L}}_{|\mathcal N|,c_1} & \cdots & \tilde{\mat{K}}_{|\mathcal N|,c_{|\mathcal N_{\text{ctrl}}|}} & \tilde{\mat{L}}_{|\mathcal N|,c_{|\mathcal N_{\text{ctrl}}|}}
\end{array}
\right]
}_{\mat{J}_{\text{ctrl}}^{-1} \in \mathbb{R}^{2|\mathcal N| \times 2|\mathcal N_{\text{ctrl}}|}}
\cdot
\underbrace{
\begin{bmatrix}
P_{\text{ctrl},c_1} z_{c_1} \\
Q_{\text{ctrl},c_1} z_{c_1} \\
\vdots \\
P_{\text{ctrl},c_{|\mathcal N_{\text{ctrl}}|}} z_{c_{|\mathcal N_{\text{ctrl}}|}} \\
Q_{\text{ctrl},c_{|\mathcal N_{\text{ctrl}}|}} z_{c_{|\mathcal N_{\text{ctrl}}|}}
\end{bmatrix}
}_{\vect{u} \in \mathbb{R}^{2|\mathcal N_{\text{ctrl}}|}}
+
\underbrace{
\begin{bmatrix}
b_1 \\
\vdots \\
b_{2|\mathcal N|}
\end{bmatrix}
}_{\vect{b} \in \mathbb{R}^{2|\mathcal N|}}
$}
\label{eq:full_block}
\end{equation}

\subsection{Volt/VAr Control}
Let the voltage deviation \(\Delta V_i^{(k)} = V_i^{(k)} - V_i^0, \quad \forall i \in \mathcal{N}\). Recall that the state vector is given by \(\vect{x}^{(k)}\) and the effective control input by \(\vect{u}^{(k)}\). In the distributed framework, \(\vect{s}_{i}^{(k)}\!\in\!\mathbb{R}^{2| \mathcal{N}_{\text{ctrl}}|}\) is the bus-level estimate of the global controllable-injection vector \(\vect{u}^{(k)}\) obtained via neighbor-only consensus (\(\vect{s}_i\!\approx\!\vect{u},\;\forall i\in\mathcal N\)). 
For each controllable bus \(i\), \(\vect{s}_{i}^{(k)}\in\mathbb{R}^{2| \mathcal{N}_{\text{ctrl}}|}\) is interpreted locally as that agent’s running estimate of \(\vect{u}^{(k)}\).

Substituting into the linearized model yields the state-space Eq.~\eqref{eq:state_space} with the last \(|\mathcal{N}|\) rows corresponding to voltage deviations. 
The control vector \(\vect{u} \in \mathbb{R}^{2|\mathcal{N}_{\text{ctrl}}|}\) stacks the effective injections (\(P_{\text{ctrl},i} z_i\), \(Q_{\text{ctrl},i} z_i\)) for all controllable buses \(i \in \mathcal{N}_{\text{ctrl}}\). Define \([\tilde{\mat{K}}_V]_i = [\tilde{\mat{K}} \, \tilde{\mat{L}}]_{i,\text{ctrl}}\), the \(i\)-th row of the voltage sensitivity matrix \([\tilde{\mat{K}} \, \tilde{\mat{L}}]_{\text{ctrl}}\) restricted to columns for \(\mathcal{N}_{\text{ctrl}}\), or equivalently, \([\tilde{\mat{K}}_V]_i = [\mat{J}_{\text{ctrl}}^{-1}]_{V_i,:}\).
The local voltage deviation at bus \(i\) is given by:
\begin{equation}
\begin{split}
\Delta V_i^{(k)} = [\tilde{\mat{K}}_V]_i \, \vect{u}^{(k)} + b_{V,i} \quad \text{(theoretical mapping)}, \\
\Delta V_i^{\text{est},(k)} \approx [\tilde{\mat{K}}_V]_i \, \vect{s}_i^{(k)} + b_{V,i}  \quad \text{(using the estimate)}
\end{split}
\label{eq:voltage_estimate_consensus}
\end{equation}
where \([\tilde{\mat{K}}_V]_i\) is the \(i\)-th row of the combined voltage-sensitivity matrices \(\tilde{\mat{K}}\) and \(\tilde{\mat{L}}\) extracted from \(\mat{J}_{\text{ctrl}}^{-1}\), mapping the global effective control vector \(\vect{u}\) to the local voltage deviation \(\Delta V_i\). \(b_{V,i}\) is the component of the offset vector \(\vect{b}\) corresponding to \(\Delta V_i\), accounting for fixed injections. 
In the distributed framework, agents maintain a local estimate $\vect{s}_i$ of the global control vector \(\vect{u}\). This allows an agent to approximate the local voltage deviation:
\begin{equation}
\Delta V_i^{\text{est}} \approx [\mat{J}_{\text{ctrl}}^{-1}]_{V_i,:} \vect{s}_i + b_{V,i},
\end{equation}
where \(\big[\mat{J}_{\text{ctrl}}^{-1}\big]_{V,:}\) denotes the block consisting of the last \(n\) rows
(voltage rows) of \(\mat{J}_{\text{ctrl}}^{-1}\) with \(n=|\mathcal N|\), and \(\big[\cdot\big]_{V_i,:}\) is its \(i\)-th row.
The actual voltage deviation \(\Delta V_i^{(k)}\) is a primal variable determined by the iterative primal gradient descent, (specifically, in the objective function term \(\alpha_i (\Delta V_i^2)\) and in the dual variable updates in Eqs.~\eqref{eq:dual_update_p} --~\eqref{eq:dual_update_q} detailed in Section~\ref{sec:primal_dual}). That iterative update directly incorporates feedback from the objective function gradient and the dual variables \((\lambda_j)\), driving the variable towards the optimal solution satisfying the problem constraints. The estimate \(\Delta V_i^{\text{est},(k)}\) -- Equation \eqref{eq:voltage_estimate_consensus} --  provides a way for the agent to gain situational awareness based on the consensus state \(\vect{s}_i^{(k)}\). 

\subsection{Network Model}
We consider a power distribution network represented by a connected graph $\mathcal{G} = (\mathcal{N}, \mathcal{E})$, where $\mathcal{N}$ is the set of buses and $\mathcal{E}$ is the set of electrical lines (edges) connecting them. Furthermore, the adjacency matrix \(\mat{A} \in \{0,1\}^{|\mathcal{N}|\times|\mathcal{N}|}\) captures the network connectivity, with \(A_{ij}=1\) if buses \(i\) and \(j\) are directly connected, and 0 otherwise. We assume that the communication network follows the same topology as the electrical network: each bus can communicate directly with its electrically connected neighbors. This means agents can exchange information with their direct electrical neighbors; i.e., they share local state variables $(\Delta V_j, \Delta \theta_j)$, dual variables $(\lambda_j^P, \lambda_j^Q)$, and the consensus state vector \(\vect{s}_j\). 

\section{Problem Formulation}\label{sec:problem_formulation}

\subsection{Variables}
Variables are defined as follows: the state variables are the voltage magnitude \(V_i^{(k)}\) and the voltage phase angle \(\theta_i^{(k)}\) at each bus \(i\) for every iteration \(k\). Voltage deviations are given by \(\Delta V_i^{(k)} = V_i^{(k)} - V_i^0 \quad \text{and} \quad \Delta \theta_i^{(k)} = \theta_i^{(k)} - \theta_i^0, \) representing the differences from their respective nominal values. For buses equipped with controllable devices, the 
control variables are the power \emph{set-points} \(P_{\text{ctrl},i}^{(k)}\) and \(Q_{\text{ctrl},i}^{(k)}\), while the activation variable \(z_i^{(k)} \in [0,1]\) acts as a \emph{dispatch factor}. The total net power injections are expressed by \(P_i^{(k)}\) for active power and \(Q_i^{(k)}\) for reactive power at each bus. The set \(\mathcal{N}\) represents all buses in the network, with \(\mathcal{N}_{\text{ctrl}}\) identifying those with controllable devices and \(\mathcal{N}_{\text{unctrl}}\) the remainder. A reference (slack) bus \(i_{\mathrm{ref}} \in \mathcal{N}\) provides the angle reference (\(\Delta \theta_{i_{\mathrm{ref}}}=0\)) and is handled distinctly in certain constraints, as detailed below (e.g., Eqs. \eqref{eq:net_p_unctrl} -- \eqref{eq:net_q_unctrl}). The set \(\mathcal{N}_i^{\text{neigh}} = \{ j \mid j \sim i \}\) defines the directly connected neighboring buses for each bus \(i\), which fully characterizes the network connectivity and topology. Nominal values, denoted with a superscript \(^0\) (e.g., \(V_i^0\), \(\theta_i^0\), \(P_i^0\), \(Q_i^0\)), represent standard operating conditions. Load demands are specified by \(P_{\text{load},i}\) and \(Q_{\text{load},i}\) for active and reactive power, respectively. The admittance matrix elements \(Y_{ij} = G_{ij} + \mathrm{j} B_{ij}\) capture the electrical connection between buses, with \(G_{ij}\) as conductance and \(B_{ij}\) as susceptance. Voltage limits are defined by \(\underline{V}\) and \(\overline{V}\). For controllable devices, limits on active and reactive power are given by \(\underline P_{\text{ctrl},i}\), \(\overline P_{\text{ctrl},i}\), \(\underline Q_{\text{ctrl},i}\), and \(\overline Q_{\text{ctrl},i}\), along with the maximum apparent power capacity \(\overline S_i\). Finally, the objective function weights \(\alpha_i\) and \(\beta_i\) are used to balance the cost associated with voltage deviations and the activation of controllable devices. We use \(k\) to index iterations and \(i\) to index buses. 
\subsection{Objective Function}
The optimization problem is:
\begin{equation}
\begin{split}
\min_{\substack{P_{\text{ctrl},i}^{(k)}, \; Q_{\text{ctrl},i}^{(k)}, \; z_i^{(k)}, \\ 
\Delta V_i^{(k)}, \; \Delta \theta_i^{(k)}}} \quad 
J^{(k)} =\;& \sum_{i \in \mathcal{N}} \alpha_i \left( \Delta V_i^{(k)} \right)^2  
+ \sum_{i \in \mathcal{N}_{\text{ctrl}}} \beta_i z_i^{(k)} \\
&+ \rho \sum_{i \in \mathcal{N}_{\text{ctrl}}} 
\phi_i\Bigl(P_{\text{ctrl},i}^{(k)},\, Q_{\text{ctrl},i}^{(k)},\, z_i^{(k)}\Bigr),
\end{split}
\label{eq:complete_problem}
\end{equation}
where the penalty function is 
\begin{equation}
\resizebox{\columnwidth}{!}{$
\phi_i\!\left(P_{\text{ctrl},i}^{(k)},Q_{\text{ctrl},i}^{(k)},z_i^{(k)}\right)
=\max\left\{0,\; \big(z_i^{(k)} P_{\text{ctrl},i}^{(k)}\big)^2 + \big(z_i^{(k)} Q_{\text{ctrl},i}^{(k)}\big)^2 - \big(\overline{S}_i z_i^{(k)}\big)^2 \right\}
$}.
\label{eq:penalty}
\end{equation}
It activates whenever $P_{\text{ctrl},i}^{(k)2}+Q_{\text{ctrl},i}^{(k)2}>\overline S_i^{\,2}$ (for any $z_i^{(k)}>0$).
The hard constraint $(z_i^{(k)} P_{\text{ctrl},i}^{(k)})^2+(z_i^{(k)} Q_{\text{ctrl},i}^{(k)})^2 \le (\overline S_i z_i^{(k)})^2$ is thus relaxed by \eqref{eq:penalty}.
For the control update, \(\partial_{z_i}\phi_i\) is taken from the subgradient of \(\phi_i\) w.r.t.\ \(z_i\) \footnote{Problem~\eqref{eq:complete_problem} is nonconvex due to the bilinear terms $P_{\text{ctrl},i}z_i$, $Q_{\text{ctrl},i}z_i$ and the penalty $\phi_i(\cdot)$; accordingly, we seek first-order stationarity (not global optimality).}. 
Let
\begin{equation}
g_i := P_{\text{ctrl},i}^2 + Q_{\text{ctrl},i}^2 - (\overline S_i)^2.
\end{equation}
Then
\begin{equation}
\partial_{z_i}\phi_i =
\begin{cases}
2\,z_i\, g_i, & g_i > 0,\\[1mm]
\{0\}, & g_i \le 0~.
\end{cases}
\label{eq:phi_subgrad}
\end{equation}

\subsection{Constraints}
\paragraph{Linearized Nodal Power Balance Equations}
For all buses \(i \in \mathcal{N}\setminus\{i_{\mathrm{ref}}\}\): 
\begin{align}
\Delta P_i^{(k)} &= \sum_{j \in \mathcal{N}} H_{ij} \Delta \theta_j^{(k)} + \sum_{j \in \mathcal{N}} N_{ij} \Delta V_j^{(k)}, \label{eq:nodal_pf_p} \\
\Delta Q_i^{(k)} &= \sum_{j \in \mathcal{N}} K_{ij} \Delta \theta_j^{(k)} + \sum_{j \in \mathcal{N}} L_{ij} \Delta V_j^{(k)}, \label{eq:nodal_pf_q}
\end{align}
where \(\Delta P_i^{(k)} = P_i^{(k)} - P_i^0, \quad \forall i \in \mathcal{N}\) and \(\Delta Q_i^{(k)} = Q_i^{(k)} - Q_i^0, \quad \forall i \in \mathcal{N}\).

\paragraph{Net Power Injections}
For buses with controllable devices (\(i \in \mathcal{N}_{\text{ctrl}} \)):
\begin{align}
P_i^{(k)} &= P_{\text{ctrl},i}^{(k)} z_i^{(k)} - P_{\text{load},i}, \label{eq:net_p_ctrl} \\
Q_i^{(k)} &= Q_{\text{ctrl},i}^{(k)} z_i^{(k)} - Q_{\text{load},i}. \label{eq:net_q_ctrl}
\end{align}
For buses without controllable devices \((i \in \mathcal{N}_{\text{unctrl}} \setminus \{ i_{\mathrm{ref}} \})\):
\begin{align}
P_i^{(k)} &= - P_{\text{load},i}, \label{eq:net_p_unctrl} \\
Q_i^{(k)} &= - Q_{\text{load},i}. \label{eq:net_q_unctrl}
\end{align}
\paragraph{Controllable Device Constraints}
For each \( i \in \mathcal{N}_{\text{ctrl}} \), the active and reactive power injections from controllable devices are subject to box constraints, with lower and upper bounds denoted by \( \underline{P}_{\text{ctrl},i}\), \( \overline{P}_{\text{ctrl},i}\) for active power, and \( \underline{Q}_{\text{ctrl},i}\), \( \overline{Q}_{\text{ctrl},i}\) for reactive power:
\begin{align} 
\underline{P}_{\text{ctrl},i}\ &\leq P_{\text{ctrl},i}^{(k)} \leq \overline{P}_{\text{ctrl},i}, \label{eq:ctrl_p_limits} \\
\underline{Q}_{\text{ctrl},i}\, &\leq Q_{\text{ctrl},i}^{(k)} \leq \overline{Q}_{\text{ctrl},i}\, \label{eq:ctrl_q_limits} \\
z_i^{(k)} &\in [0, 1]. \label{eq:activation_var}
\end{align}
These box constraints ensure the \emph{set-points} remain within the device's \emph{fixed} hardware limits
\cite{ieee_ieee_2018}, where we relax the binary constraint \(z_i^{(k)} \in \{0,1\}\) to \(z_i^{(k)} \in [0, 1]\). We consider setting limits with both negative and positive values to represent a device that can both inject and absorb power.
\begin{align}
\underline{P}_{\text{ctrl},i} = \underline{Q}_{\text{ctrl},i} &= -\overline{S}_i, \\ 
\overline{P}_{\text{ctrl},i} = \overline{Q}_{\text{ctrl},i} &= +\overline{S}_i. 
\end{align}
creating a symmetrical operating range that allows for four-quadrant operation. The device can be a source or a sink for both active and reactive power, enabling it to operate in any of the four P-Q quadrants. 
The overarching apparent-power capability $\overline S_i$ defines the disk $P^2+Q^2\le \overline S_i^2$ in the $(P,Q)$ plane. In our formulation this circular limit is enforced \emph{softly} via the penalty $\phi_i$ (rather than as a hard constraint), so the hard feasible set is the rectangle given by the box limits, and violations of the disk are discouraged by the penalty term.
\paragraph{Voltage Constraints}
For all \(i \in \mathcal{N}\):
\begin{equation}
\underline{V} \leq V_i^0 + \Delta V_i^{(k)} \leq 
\overline{V}.
\label{eq:voltage_limits}
\end{equation}
\paragraph{Voltage Angle Constraints} To ensure stability, the angle deviation at each non-reference bus is limited:
\begin{equation}
|\Delta\theta_i^{(k)}| \le \frac{\pi}{6}, \quad \forall i \in \mathcal{N} \setminus \{i_{\mathrm{ref}}\}.
\label{eq:angle_limit}
\end{equation}
The reference bus angle and voltage are fixed: \(\Delta \theta_{i_{\mathrm{ref}}}=0,\quad \Delta V_{i_{\mathrm{ref}}}=0.\)
\subsection{Decision Variables}
The optimization variables are \(\{\Delta V_i^{(k)},\,\Delta \theta_i^{(k)}\}_{i\in\mathcal N}\) and \(\{P_{\text{ctrl},i}^{(k)},\,Q_{\text{ctrl},i}^{(k)},\,z_i^{(k)}\}_{i\in\mathcal N_{\text{ctrl}}}\).
The goal is to determine \(P_{\text{ctrl},i}^{(k)}\), \(Q_{\text{ctrl},i}^{(k)}\), \(z_i^{(k)}\) together with \(\Delta V_i^{(k)}\), \(\Delta \theta_i^{(k)}\) (Recover \(V_i^{(k)}=V_i^0+\Delta V_i^{(k)}\), \(\theta_i^{(k)}=\theta_i^0+\Delta \theta_i^{(k)}\) when needed).
The linearization around the operating point \((V^0,\theta^0)\) ensures that the power flow equations capture the effect of variations in voltage magnitude and phase angle — an important feature in networks with high distributed generation and bidirectional power flows. The penalty formulation for the apparent power constraints helps maintain the operation of controllable devices within their physical limits, namely when adjusting control actions during method iterations. 
\section{Decentralized Saddle-Point Reformulation} \label{sec:methodology}
Two architectures are commonly used for distributed Volt/VAr control. In \textit{zone-based hierarchical control}, the network is partitioned into zones (such as VPPs or microgrids), each with a local slack bus and centralized controller \cite{pierrou_decentralized_2024}. In contrast, \textit{fully decentralized agent-based control} assigns an autonomous agent to each controllable bus. These agents update their own control and dual variables using only local measurements and communication with neighboring buses \cite{wu_enhanced_2016}. Without requiring any centralized coordinator, the agents collectively solve the system-wide coordination problem through consensus-based estimation and distributed saddle-point updates. Building on the optimization problem presented in Section ~\ref{sec:problem_formulation}, we reformulate the centralized problem as a distributed saddle-point problem \cite{alghunaim_dual_2021} by dualizing the coupling (power balance) constraints. We adopt the fully decentralized agent-based formulation, where agents at controllable buses \(i \in \mathcal{N}_{\text{ctrl}}\) and non-controllable buses \(i \in \mathcal{N} \setminus \mathcal{N}_{\text{ctrl}}\) participate in the optimization to obtain a feasible solution.
\subsection{Saddle-Point Reformulation}\label{sec:decentralized_saddle}
We aim to solve the original centralized problem in a distributed manner using a saddle-point method based on the global Lagrangian \eqref{eq:lagrangian_full}. 
Coordination is achieved via dual variable updates and via C+I to track shared estimates. We dualize the coupling constraints—the linearized active/reactive power-balance equations~\eqref{eq:nodal_pf_p} -– \eqref{eq:nodal_pf_q}—to form the Lagrangian. Let \(\vect{x} = \{ \Delta V_i, \Delta \theta_i \}_{i \in \mathcal{N}}\) be the collection of all state variables, \(\vect{u} = \{ P_{\text{ctrl},i}, Q_{\text{ctrl},i}, z_i \}_{i \in \mathcal{N}_{\text{ctrl}}}\) be the control variables, and \(\vect{\lambda} = \{ \lambda_i^P, \lambda_i^Q \}_{i \in \mathcal{N}\setminus\{i_{\mathrm{ref}}\}}\) be the dual variables (Lagrange multipliers) associated with the active and reactive power balance constraints.
We use a decentralized primal–dual heuristic to seek a first-order stationary point:
\begin{equation}
\min_{\vect{x}\in\mathcal{X},\,\vect{u}\in\mathcal{U}}\;
\max_{\vect{\lambda}\in\Lambda}\;
\mathcal{L}(\vect{x},\vect{u},\vect{\lambda}),
\label{eq:sp_formulation}
\end{equation} 
with $\Lambda=\mathbb{R}^{2(|\mathcal{N}|-1)}.$ 
\begin{equation}
\begin{aligned} 
\mathcal{L}(\vect{x},\vect{u},\vect{\lambda}) =
&\sum_{i\in\mathcal N} \alpha_i(\Delta V_i)^2 \\
& + \sum_{i \in \mathcal{N}\setminus\{i_{\mathrm{ref}}\}}
\Big[
  \lambda_i^P\big(\Delta P_i - \sum_j H_{ij}\Delta\theta_j - \sum_j N_{ij}\Delta V_j\big) \\
&  + \lambda_i^Q\big(\Delta Q_i - \sum_j K_{ij}\Delta\theta_j - \sum_j L_{ij}\Delta V_j\big)
\Big] \\
& + \sum_{i\in\mathcal N_{\text{ctrl}}}\beta_i z_i 
+ \rho\sum_{i\in\mathcal N_{\text{ctrl}}}\phi_i(P_{\text{ctrl},i},\, Q_{\text{ctrl},i},\, z_i).
\end{aligned}
\label{eq:lagrangian_full}
\end{equation}

The Lagrangian \(\mathcal{L}(\vect{x},\vect{u}, \vect{\lambda})\) includes both smooth and nonsmooth components of the objective, including the penalty term \(\rho\,\phi_i(\cdot)\) for each bus \(i \in \mathcal{N}_{\text{ctrl}}\), and the dual domain is \(\Lambda = \mathbb{R}^{2(|\mathcal{N}|-1)}\).  At the saddle point \((\vect{x}^*, \vect{u}^*,\vect{\lambda}^*)\), the following condition holds for all \(\vect{x}\in\mathcal{X}\), \(\vect{u}\in\mathcal{U}\) and \(\vect{\lambda}\in\Lambda\): 
\begin{equation}
\begin{split}
\mathcal{L}(\vect{x}^*,\vect{u}^*,\vect{\lambda})
\le \mathcal{L}(\vect{x}^*,\vect{u}^*,\vect{\lambda}^*) 
\le \mathcal{L}(\vect{x},\vect{u},\vect{\lambda}^*),\\
\forall\,\vect{x}\in\mathcal{X},\;
      \vect{u}\in\mathcal{U},\;
      \vect{\lambda}\in\Lambda. 
\end{split}
\label{eq:saddle_point}
\end{equation}

where \(\mathcal{X}\) is the set of all primal variables satisfying the local constraints (device limits, voltage limits, reference bus constraint). \(\mathcal{U}\) is the set of all primal variables satisfying the local control constraints. \(\Lambda\) is the (unconstrained) space of dual variables. The constraint sets are given by \(\mathcal{X} = \Bigl\{ \{\Delta V_i, \Delta \theta_i\}_{i\in\mathcal{N}} \mid \underline{V} \le V_i^0 + \Delta V_i \le \overline{V}, \forall i \in \mathcal{N}; |\Delta \theta_i| \le \pi/6, \forall i \in \mathcal{N}\setminus\{i_{\mathrm{ref}}\}; \Delta\theta_{i_{\mathrm{ref}}}=0, \Delta V_{i_{\mathrm{ref}}}=0 \Bigr\}\) and
\(\mathcal{U} = \Bigl\{ P_{\text{ctrl},i}, Q_{\text{ctrl},i}, z_i : \underline{P}_{\text{ctrl},i} \le P_{\text{ctrl},i} \le \overline{P}_{\text{ctrl},i},\; \underline{Q}_{\text{ctrl},i} \le Q_{\text{ctrl},i} \le \overline{Q}_{\text{ctrl},i},\; z_i \in [0,1] \Bigr\}\)\footnote{Note that \(\mathcal{U}\) excludes the apparent power constraint, which is handled by the penalty function \(\phi_i\) in the objective function \eqref{eq:penalty}.}. 
The set \(\mathcal{U}\) is defined using only linear inequalities; the non-linear circular constraint on apparent power is handled separately by the penalty term in the objective function. Let the local objective at bus \(i\) be denoted by \(J_i\). We split the local objective at bus \(i\) into a smooth component (\(f(x_i,u_i)\)) and a nonsmooth \(\phi(\vect{u}_i)\) penalty (e.g., enforcing apparent power limits) that depends only on the control variables: \(J_i(\vect{x}_i,\vect{u}_i) = f(\vect{x}_i,\vect{u}_i) + \rho\, \phi_i(\vect{u}_i)\).
Each bus also updates its dual variables via gradient ascent to enforce the coupling constraints, which correspond to local power balance equations, ensuring system-wide feasibility without requiring full system information at each node. This decomposition reflects the structure of the decentralized updates: gradient descent is applied to the smooth term \(f_i\), while the nonsmooth penalty \(\phi_i\) is handled via a proximal operator on the control variables \(\vect{u}_i\).
\subsection{Distributed Primal-Dual Updates}
\label{sec:primal_dual}
The saddle-point problem is solved iteratively using a distributed primal-dual method. Each bus \(i \in \mathcal{N}\) maintains and updates its local variables based on local information and communication with its neighbors \(j\in\mathcal{N}_i^{\text{neigh}}\). The algorithm alternates between primal descent steps to minimize the Lagrangian with respect to primal variables \((\vect{x}, \vect{u})\) and a dual ascent step to maximize it with respect to dual variables \((\vect{\lambda})\). The local Jacobian neighborhood is defined as \(\mathcal N_i^{J}:=\{\,j\mid H_{ij}\neq0\ \lor\ N_{ij}\neq0\ \lor\ K_{ij}\neq0\ \lor\ L_{ij}\neq0\ \lor\ H_{ji}\neq0\ \lor\ N_{ji}\neq0\ \lor\ K_{ji}\neq0\ \lor\ L_{ji}\neq0\,\}\).
For typical power systems, the Jacobian matrix \(\mat{J}\) is sparse, so $\mathcal {N}_i^{J}$ contains only bus \(i\) and its immediate neighbors, enabling neighbor-only communication.
\paragraph{Primal Update: State Variables \((\Delta V_i, \Delta \theta_i)\)}
All buses $i \in \mathcal{N} \setminus \{i_{\mathrm{ref}}\}$ update their state variables via projected gradient descent. The gradients are computed locally using shared dual variables from neighbors:
\begin{align}
\nabla_{\Delta V_i} \mathcal{L} &= 2\alpha_i \Delta V_i - \sum_{j \in \mathcal{N}_i^{J}}  (\lambda_j^P N_{ji} + \lambda_j^Q L_{ji}), \\
\nabla_{\Delta \theta_i} \mathcal{L} &= - \sum_{j \in \mathcal{N}_i^{J}}  (\lambda_j^P H_{ji} + \lambda_j^Q K_{ji}).
\end{align}
The update steps, with step-size $\eta_1$, involve projection onto the feasible sets defined by \eqref{eq:voltage_limits} and \eqref{eq:angle_limit}:
\begin{align}
\Delta V_i^{(k+1)} &\leftarrow \Pi_{[\underline{V}-V_i^0,\, \overline{V}-V_i^0]} \left[ \Delta V_i^{(k)} - \eta_1 \nabla_{\Delta V_i} \mathcal{L}^{(k)} \right], 
\label{eq:dV_update_final} \\
\Delta \theta_i^{(k+1)} &\leftarrow \Pi_{[-\pi/6, \,\pi/6]} \left[ \Delta \theta_i^{(k)} - \eta_1 \nabla_{\Delta \theta_i} \mathcal{L}^{(k)} \right]. \label{eq:dTheta_update_final}
\end{align}

\paragraph{Primal Update: Control Variables \(\vect{u}_i\)}
For each controllable bus \(i \in \mathcal{N}_{\text{ctrl}}\), the control variables are updated using a two-stage Gauss-Seidel scheme. 
First, the activation variable $z_i$ is updated via a projected gradient descent step on the Lagrangian. With $\partial_{z_i}\mathcal{L}^{(k)} = \beta_i + \lambda_i^{P,(k)} P_{\text{ctrl},i}^{(k)} + \lambda_i^{Q,(k)} Q_{\text{ctrl},i}^{(k)} + \rho\,\partial_{z_i}\phi_i$, the update is:
\begin{equation}
z_i^{(k+1)} \gets \Pi_{[0,1]}\!\left( z_i^{(k)} - \eta_2 \partial_{z_i}\mathcal{L}^{(k)} \right).
\end{equation}
Second, using this updated \(z_i^{(k+1)}\), the power set-points are updated via a proximal gradient step. This involves a gradient descent on the smooth part of the Lagrangian, followed by application of the proximal operator for the non-smooth penalty \(\phi_i\) (for the closed-form proximal operator)\footnote{Let $v=(P',Q')^\top$. The proximal operator of $\eta\rho\,\phi_i(\cdot,\cdot,z)$ is radial: $w^*=\kappa v$ with $\kappa=\min\{1,\max\{\overline S_i/\|v\|_2,\ 1/(1+2\eta\rho z^2)\}\}$.  Case $\|v\|\le \overline S_i$: no penalty, $\kappa=1$. Otherwise, the shrink solution $\kappa=1/(1+2\eta\rho z^2)$ competes with the boundary point $\kappa=\overline S_i/\|v\|$.}, and a final projection onto \(\mathcal{U}_i\): 
\begin{align} 
    \begin{bsmallmatrix} P' \\ Q' \end{bsmallmatrix} &\gets
    \begin{bsmallmatrix} P_{\text{ctrl},i}^{(k)} \\ Q_{\text{ctrl},i}^{(k)} \end{bsmallmatrix}
    - \eta_2
    \begin{bsmallmatrix} \lambda_i^{P,(k)} z_i^{(k+1)} \\ \lambda_i^{Q,(k)} z_i^{(k+1)} \end{bsmallmatrix}
    \label{eq:prox_grad_step}, \\
    \begin{bsmallmatrix} P'' \\ Q'' \end{bsmallmatrix} &\gets
    \operatorname{prox}_{\eta_2 \rho \phi_i}\!\left(
    \begin{bsmallmatrix} P' \\ Q' \end{bsmallmatrix}
    \right),
    \label{eq:prox_op_step} \\    
    \begin{bsmallmatrix} P_{\text{ctrl},i}^{(k+1)} \\ Q_{\text{ctrl},i}^{(k+1)} \end{bsmallmatrix} 
    &\gets
    \Pi_{\mathcal{U}_i}\!\left(
    \begin{bsmallmatrix} P'' \\ Q'' \end{bsmallmatrix}
    \right). 
    \label{eq:proj_step}
\end{align} 

\paragraph{Projection onto $\mathcal{U}_i$} 
Define
\begin{equation}
\begin{split}
\mathcal{U}_i:=\{(P,Q)\in\mathbb{R}^2 \mid\;
& \underline{P}_{\text{ctrl},i}  \le P \le \overline{P}_{\text{ctrl},i} ,\\
& \underline{Q}_{\text{ctrl},i}  \le Q \le \overline{Q}_{\text{ctrl},i}\}
\end{split}
\label{eq:U_set_def}
\end{equation}

The Euclidean projection onto \(\mathcal{U}_i\) is computed by clipping \((P'',Q'')\) coordinate‐wise to the box \([\underline P_{\text{ctrl},i},\overline P_{\text{ctrl},i}]\times[\underline Q_{\text{ctrl},i},\overline Q_{\text{ctrl},i}]\). 
\paragraph{Dual Update: Lagrange Multipliers \((\lambda_i)\)}
All buses update their dual variables via gradient ascent on the Lagrangian. The gradient is the primal residual, which measures the mismatch in the power balance equations using the latest state and control variables \((\vect{x}^{(k+1)}, \vect{u}^{(k+1)})\) from the primal update. The update equations are \(\forall\, i\in\mathcal N\setminus\{i_{\mathrm{ref}}\}\):
\begin{align}
\lambda_i^{P,(k+1)} &\leftarrow \lambda_i^{P,(k)} + \gamma \bigg( \underbrace{P_{\text{ctrl},i}^{(k+1)} z_i^{(k+1)}}_{\text{0 if } i \notin \mathcal{N}_{\text{ctrl}}} - P_{\text{load},i} - P_i^0 \nonumber\\
&\quad - \sum_{j \in \mathcal{N}_i^{J}} \big( H_{ij}\Delta \theta_j^{(k+1)} + N_{ij}\Delta V_j^{(k+1)} \big) \bigg), \label{eq:dual_update_p} \\[2mm]
\lambda_i^{Q,(k+1)} &\leftarrow \lambda_i^{Q,(k)} + \gamma \bigg( \underbrace{Q_{\text{ctrl},i}^{(k+1)} z_i^{(k+1)}}_{\text{0 if } i \notin \mathcal{N}_{\text{ctrl}}} - Q_{\text{load},i} - Q_i^0 \nonumber\\
&\quad - \sum_{j \in \mathcal{N}_i^{J}} \big( K_{ij}\Delta \theta_j^{(k+1)} + L_{ij}\Delta V_j^{(k+1)} \big) \bigg), \label{eq:dual_update_q}
\end{align} 
where \(\lambda_{i_{\mathrm{ref}}}^{P}=\lambda_{i_{\mathrm{ref}}}^{Q}=0,\) and $\gamma > 0$ is the dual step-size, and \(\mathcal{N}_i^{J}\) is as defined above.

\subsection{Coordination via C+I}
A challenge in a decentralized setting is ensuring that local control actions are globally coordinated. We achieve this using a C+I mechanism. Each agent
\(i\) in the entire network \(\forall i \in \mathcal{N}\) maintains a local estimate, \(\vect{s}_i \in \mathbb{R}^{2\lvert \mathcal{N}_{\text{ctrl}}\rvert}\) of the global vector of effective injections \(\vect{u}\). Agents at non-controllable buses \(i \notin \mathcal{N}_{\text{ctrl}}\) participate in the consensus protocol by relaying information from their neighbors, but their local innovation term is zero. This ensures that the consensus state \(\vect{s}\) is correctly propagated across the entire communication graph, which mirrors the electrical network \(\mathcal{G}\). Through an inner loop of communication with its neighbors, each agent updates its estimate \(\vect{s}_i\) to align with others, while also incorporating its own local control proposal \(\vect{\hat{u}}_i = [P_{\text{ctrl}, i}z_i, Q_{\text{ctrl}, i}z_i]^\top\). The local control vector is defined in Eq.~\eqref{eq:u_and_u_hat_def}. 
The update for agent \(i\)'s estimate \(\vect{s}_i\) at inner consensus iteration \(t\) is:
\begin{equation}
\resizebox{\columnwidth}{!}{$
\vect{s}_i^{(t+1)} \leftarrow
\underbrace{\vect{s}_i^{(t)}
            -\alpha_{\text{con}}\sum_{j\in \mathcal{N}_i^{\text{neigh}}}\big(\vect{s}_i^{(t)}-\vect{s}_j^{(t)}\big)
            }_{\text{Consensus term (diffusion)}}
+\underbrace{\;\mathbf{1}_{\{\,i\in\mathcal N_{\text{ctrl}}\,\}}\;\mat{E}_i\bigl(\hat{\vect{u}}_i^{(k)}
           -\mat{E}_i^{\top}\vect{s}_i^{(t)}\bigr)\vphantom{\sum_{j\in\mathcal{N}_i}}}_{\text{Innovation term (local input)}}
$}   
\label{eq:consensus_unified}
\end{equation}
where $0<\alpha_{\text{con}}<2/\lambda_{\max}(\mat{L}^{\text{graph}})$ with $\mat{L}^{\text{graph}}$ the (symmetric) graph Laplacian. The diffusion map is a strict contraction on the disagreement subspace $\mathbf{1}^\perp \otimes \mathbb{R}^{2n_c}$. Equivalently,\(\mat{E}_i\bigl(\hat{\vect{u}}_i - \mat{E}_i^\top\vect{s}_i\bigr) = \mat{E}_i\hat{\vect{u}}_i - (\mat{E}_i\mat{E}_i^\top)\,\vect{s}_i,\) where \(\mat{E}_i\mat{E}_i^\top\) projects onto the \(i\)-th  \((P,Q)\) block (rows \(2p(i){-}1,\,2p(i)\)) of \(\mathbb{R}^{2n_c}\).
Let \(n_c=|\mathcal N_{\text{ctrl}}|\) and stack \(\vect{S}=\operatorname{col}(\vect s_1,\ldots,\vect s_{|\mathcal N|})\in\mathbb{R}^{2 n_c |\mathcal N|}\) with each \(\vect{s}_i\in\mathbb{R}^{2 n_c}\).
To act simultaneously on active and reactive channels, we form the block Laplacian:
\begin{equation}
  \mat{L}_{\mathrm{big}} := \mat{L}^{\mathrm{graph}} \otimes \mat{I}_{2 n_c}
  \in \mathbb{R}^{(2 n_c |\mathcal N|)\times(2 n_c |\mathcal N|)}.
\label{eq:blockL}
\end{equation}
where “\(\otimes\)” denotes the Kronecker product. The block Laplacian \(\mat{L}_{\mathrm{big}}\) operates on the stacked consensus state \(\vect{S} = \operatorname{col}(\vect{s}_1, \dots, \vect{s}_{|\mathcal{N}|})\), where each \(\vect{s}_i \in \mathbb{R}^{2 n_c}\), ensuring coordinated updates across all buses. Each contiguous block of \(2|\mathcal{N}_{\text{ctrl}}|\) rows corresponds to bus \(i\)’s estimate \(\vect{s}_i\), so \(\mat{L}_{\mathrm{big}}\) updates them in lock‑step during consensus. In this formulation, \(\mat{L}^{\text{graph}}\) is the graph Laplacian matrix, defined as  \(\mat{L}^{\text{graph}} = \mat{D} - \mat{A}\), where \(\mat{D}\) is the diagonal matrix of node degrees and \(\mat{A}\) is the adjacency matrix of the communication graph. This structure ensures the update is local, sparse, and respects vector alignment, making the C+I method both communication-efficient and structurally compatible with the decentralized Volt/VAr control task. Each bus \(i\) uses the subset of \(\vect{s}_i\) corresponding to relevant controllable injections when computing its local voltage deviation. The consensus estimate \(\vect{s}_i\) serves as a coordinator, enabling each bus to act based on a shared understanding of the desired global control state (e.g. \textit{``distributed restricted agreement problem''}\cite{hug_consensus_2015}). The innovation term ensures that local control actions (\(\vect{\hat{u}}_i^{(k)}\)) contribute to refining the global estimate \(\vect{s}_i\), indirectly affecting the system state.
\section{Numerical Simulations}
We validate the proposed decentralized control algorithm on a modified version of the CIGRE 44-bus low-voltage radial distribution network \cite{cigre_benchmark_2014}, depicted in Fig.~\ref{fig:cigre}. All simulations were implemented using the \texttt{pandapower} library in Python \cite{thurner_pandapoweropen-source_2018} \footnote{The code is available at \url{https://github.com/d-vf/Var_Volt_Control}}. To create a scenario with significant voltage stress, all loads were modeled as constant power devices drawing 55 kVA at a 0.7 lagging power factor. Seven buses were equipped with controllable static generators (\textit{sgen}), each with a nominal apparent power capacity of 70 kVA, to provide voltage support. Furthermore, to induce a localized voltage sag, the reactive power demand at bus 18 was doubled. These conditions create a challenging environment with low voltages, particularly at the ends of the feeder, requiring active intervention. 
{\setlength{\abovecaptionskip}{0pt}%
 \setlength{\belowcaptionskip}{0pt}%
 \setlength{\textfloatsep}{0pt}%
 \setlength{\floatsep}{0pt}%
\begin{figure}[!t]
  \centering
  \caption{Network and final controllable-bus set-points.}
  \subfloat[Adapted CIGRE 44-bus low-voltage radial distribution network.\label{fig:cigre}]{
    \includegraphics[width=.47\columnwidth]{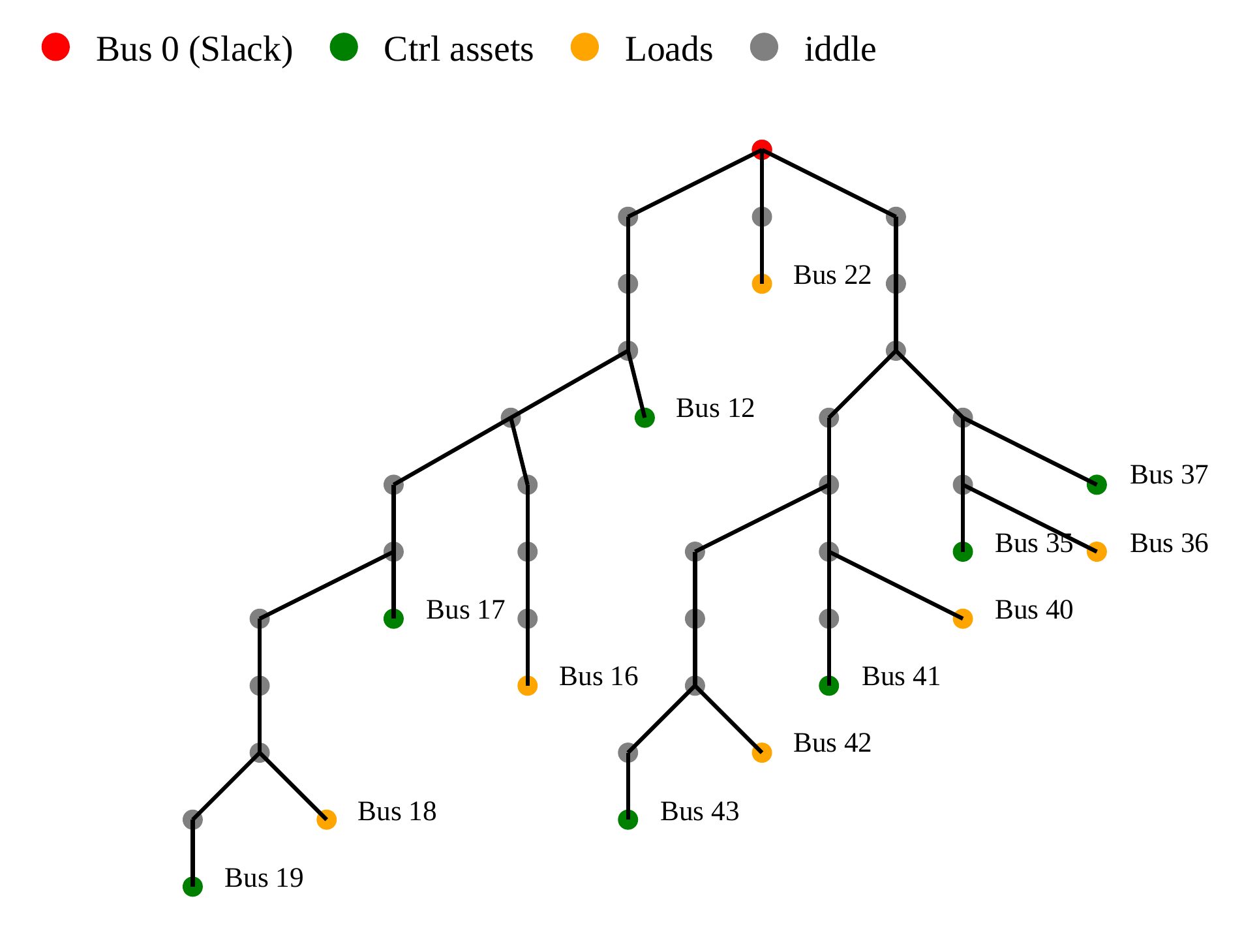}}
  \hspace{\fill}
  \subfloat[Final Controllable Bus set-points.\label{tab:final_setpoints}]{
    \begin{minipage}[t]{.47\columnwidth}\centering
      \raisebox{40pt}[20pt][00pt]{%
        \begin{adjustbox}{width=\linewidth,center}
          \small\setlength{\tabcolsep}{4pt}
          \begin{tabular}{c c c c}
            \toprule
            \textbf{Bus} & \textbf{\(P\) (kW)} & \textbf{\(Q\) (kVAr)} & \textbf{\(z\)} \\
            \midrule
            12 & 48.1 & 50.9 & 0.891 \\
            17 & 47.1 & 51.8 & 0.891 \\
            19 & 44.8 & 54.0 & 0.893 \\
            35 & 48.1 & 50.7 & 0.890 \\
            37 & 48.5 & 50.4 & 0.890 \\
            41 & 48.7 & 50.2 & 0.890 \\
            43 & 48.7 & 50.2 & 0.890 \\
            \bottomrule
          \end{tabular}
        \end{adjustbox}}
    \end{minipage}}
\end{figure}
}
{\setlength{\abovecaptionskip}{-2pt}%
 \setlength{\belowcaptionskip}{-4pt}%
 \setlength{\textfloatsep}{0pt}%
  \setlength{\floatsep}{0pt}%
\begin{figure}[!t]
  \centering
  \vspace*{-4pt}
  \caption{Voltage drop — intervention vs.\ non-intervention.}
  \vspace*{-4pt}
  \subfloat[Baseline voltage profile without intervention.]{
    \includegraphics[width=.47\columnwidth]
    {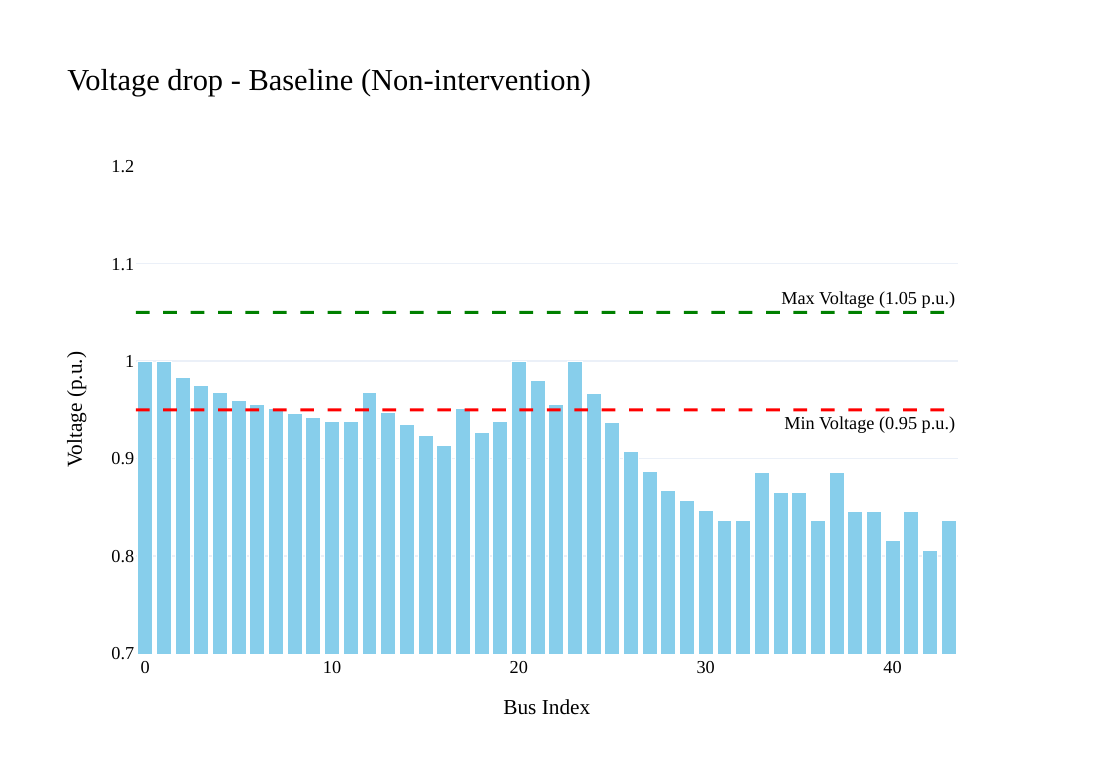}}
    \vspace*{-4pt} 
  \hspace{\fill}
  \subfloat[Voltage profile after control intervention.]{
    \includegraphics[width=.47\columnwidth]{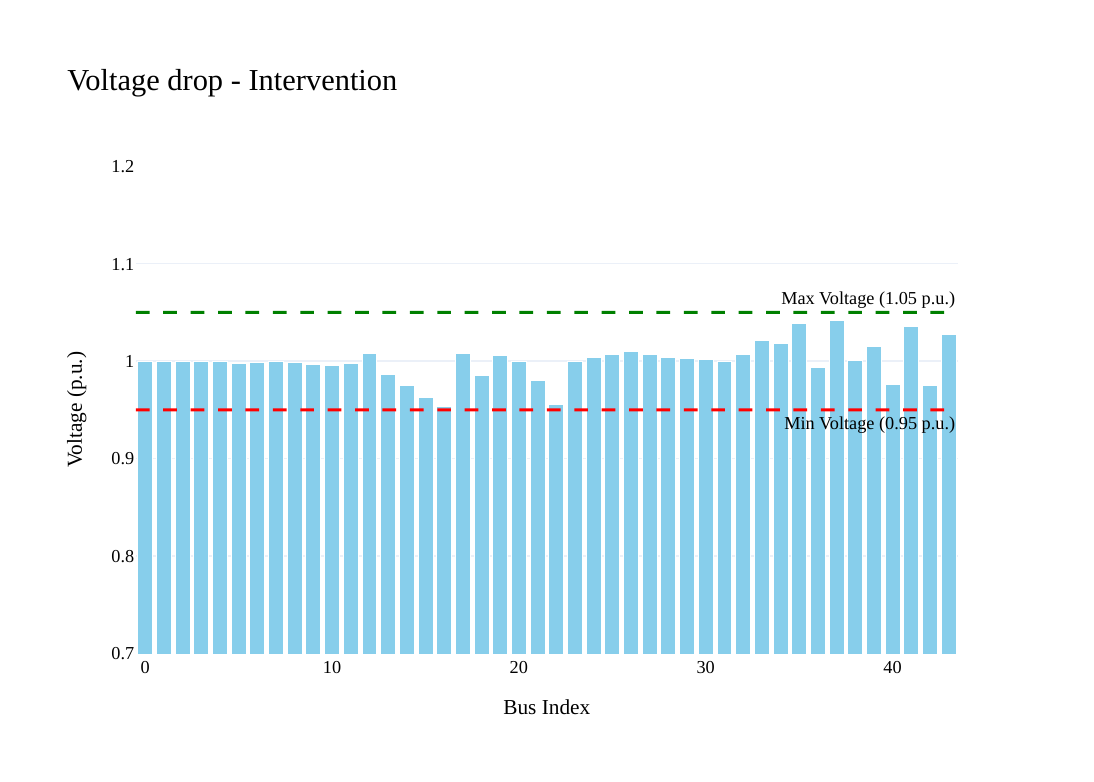}} 
  \label{fig:voltage_profiles}
\end{figure}
}
\subsection{Control Performance and Voltage Regulation}
The controller's primary objective is to mitigate voltage violations. Figure~\ref{fig:voltage_profiles} compares the voltage profiles across the network before and after the decentralized control intervention. In the baseline case (uncontrolled), voltages at several buses drop below the lower limit of 0.95 p.u., with the lowest voltage reaching 0.80 p.u. (bus 42). After the control algorithm converged, the agents' coordinated active and reactive power injections successfully raised the minimum voltage to 0.95 p.u., resolving all undervoltage issues. Table~\ref{tab:final_setpoints} details the final steady-state set-points for each of the seven controllable assets. The algorithm utilized all available resources, activating each device with $z_i \approx 0.89$ and dispatching a mix of active and reactive power to support the grid. 
The method converges empirically to a first-order stationary point; no claim of global optimality is made. With parameters set as follows \(\alpha=1\) (voltage deviation weight), \(\beta=0.01\) (control cost), \(\rho=0.1\) (penalty weight), step-sizes \(\eta_{1}=\eta_{2}=5\times10^{-3}\) (state and control, respectively) and \(\gamma=5\times10^{-5}\) (dual update), the stationarity gap steadily decreased, satisfying the convergence tolerance of $\varepsilon_{\text{iterate}}=5\times10^{-4}$ at iteration 2167.
\subsection{Linear Model Validation}
To validate its accuracy, we compared the voltage profile predicted by the controller with the results of a full, nonlinear AC power flow (AC--PF) using the final dispatched set-points. Table~\ref{tab:validation_error_abridged} provides a per-bus analysis of the prediction error. The overall Root Mean Square Error (RMSE) between the predicted and validated voltages is  0.0108 p.u., confirming a high degree of accuracy for the linear model. However, the analysis also highlights the model's limitations. Large errors occur at buses 15, 16, 22, 35, 37, 41, and 43 (e.g., 0.0379 p.u. at bus 15, and -0.0365 p.u. at bus 37), often at non-controllable buses. This is an expected outcome when large control actions shift the system's operating point away from the initial point of linearization. The decentralized control algorithm significantly reduced voltage deviations under intervention, particularly at end-of-line buses (Fig.~\ref{fig:voltage_profiles}). 
{\setlength{\abovecaptionskip}{-2pt}%
 \setlength{\belowcaptionskip}{-4pt}%
 \setlength{\textfloatsep}{0pt}%
\begin{figure}[!t]
  \centering
   \caption{Validation against AC--PF.}
  \subfloat[Predicted vs.\ AC--PF validation.\label{fig:validation_comparison}]{
    \includegraphics[width=.47\columnwidth]{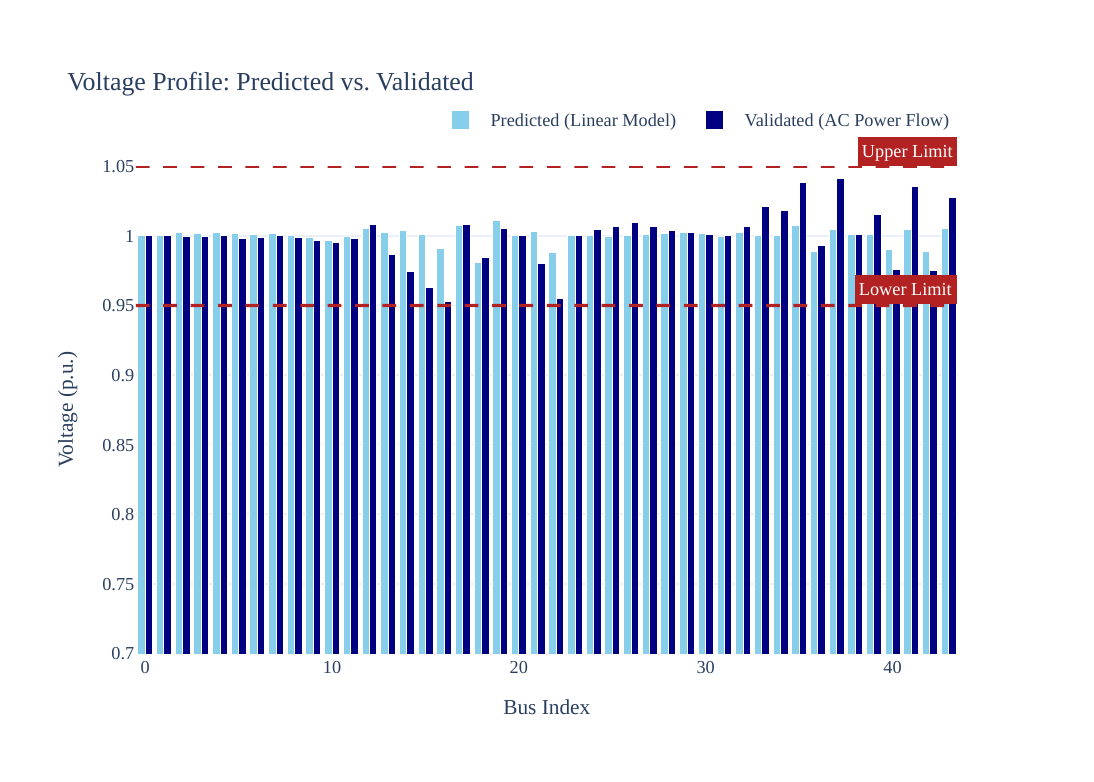}}
  \hspace{\fill}
  \subfloat[Per-bus voltage prediction error (abridged).\label{tab:validation_error_abridged}]{
    \begin{minipage}[t]{.47\columnwidth}\centering
      \raisebox{40pt}[20pt][00pt]{%
        \begin{adjustbox}{width=\linewidth,center}
          \scriptsize\setlength{\tabcolsep}{3.0pt}
          \begin{tabular}{c c c c c}
            \toprule
            \textbf{Bus} & \textbf{Predicted V} & \textbf{Validated V} & \textbf{Signed Error} & \textbf{Relative Error} \\
            & \textbf{(p.u.)} & \textbf{(p.u.)} & \textbf{(p.u.)} & \textbf{(\%)} \\ \midrule    
              0  & 1.000 & 1.000 &  0.000 &  0.00 \\
              12 & 1.005 & 1.008 & -0.002 & -0.29 \\
              15 & 1.001 & 0.963 & +0.037 & +3.94 \\
              16 & 0.990 & 0.953 & +0.037 & +3.95 \\
              19 & 1.010 & 1.005 & +0.005 & +0.51 \\
              35 & 1.007 & 1.038 & -0.031 & -3.00 \\
              37 & 1.004 & 1.041 & -0.036 & -3.51 \\
              38 & 1.000 & 1.000 &  0.000 &  0.00 \\
            \bottomrule
          \end{tabular}
        \end{adjustbox}}
    \end{minipage}}
\end{figure}
} 
\section{Conclusion}
This work presented a fully decentralized Volt/VAr control scheme for low-voltage distribution networks, based on a saddle-point formulation and local control updates. The method enables distributed assets to autonomously regulate voltage using only neighbor communication and local measurements. Numerical results demonstrated that the approach effectively mitigates voltage deviations under stress scenarios, without requiring centralized coordination. 
\bibliography{references_1}
\bibliographystyle{ieeetr}

\end{document}